\documentclass[%
 aip,
 amsmath,amssymb,
 reprint,%
]{revtex4-1}

\usepackage{graphicx}% Include figure files
\usepackage{dcolumn}% Align table columns on decimal point
\usepackage{bm}% bold math
\usepackage[utf8]{inputenc}
\usepackage[T1]{fontenc}
\usepackage{mathptmx}
\usepackage{siunitx}

\begin{document}

\preprint{AIP/123-QED}

\title{Engineering quantum-coherent defects: the role of substrate miscut in chemical vapor deposition diamond growth}

\author{S. A. Meynell}
 \email{simonmeynell@physics.ucsb.edu}
 \affiliation{Physics Department, University of California, Santa Barbara.}
 
\author{C. A. McLellan}
 \affiliation{Department of Materials Science and Engineering, Stanford University.}

\author{L. B. Hughes}
\affiliation{Materials Department, University of California, Santa Barbara.}

\author{T. E. Mates}
\affiliation{Materials Department, University of California, Santa Barbara.}

\author{K. Mukherjee}
\affiliation{Materials Department, University of California, Santa Barbara.}

\author{A. C. Bleszynski Jayich}
\affiliation{Physics Department, University of California, Santa Barbara.}

\date{\today}

\begin{abstract}
The engineering of defects in diamond, particularly nitrogen-vacancy (NV) centers, is important for many applications in quantum science. 
A materials science approach based on chemical vapor deposition (CVD) growth of diamond and in-situ nitrogen doping is a promising path toward tuning and optimizing the desired properties of the embedded defects. 
Herein, with the coherence of the embedded defects in mind, we explore the effects of substrate miscut on the diamond growth rate, nitrogen density, and hillock defect density, and we report an optimal angle range between $\SI{0.66}{\degree}< \theta<\SI{1.16}{\degree}$ for the purposes of engineering coherent ensembles of NV centers in diamond.
We provide a model that quantitatively describes hillock nucleation in the step-flow regime of CVD growth, shedding insight on the physics of hillock formation.
We also report significantly enhanced incorporation of nitrogen at hillock defects, opening the possibility for templating hillock-defect-localized NV center ensembles for quantum applications.
\end{abstract}

\maketitle

Point defects in the solid state are a promising platform for quantum science and technology.\cite{Awschalom2013,Steger2012,awschalom2018quantum} 
Diamond is a particularly attractive host material because of its high Debye temperature, wide band gap, and deep level defects, allowing for the investigation of its defect qubits up to and above room temperature.\cite{toyli2012above600K_NV}
Engineering the local environment of the qubits, both the bulk material and proximal surface, is crucial to their performance, affecting quantum coherence, \cite{ethiermajcher2017, sohn2018controlling, myers2014surfacenoise} charge state, \cite{bluvstein2019chargeinstabilities} and indistinguishability of qubits.\cite{chu2014,chu2015quantum}
Furthermore, controlling the defect density and distribution while maintaining material quality is important; for example, dense and homogeneous defect layers are required for spin ensemble-based sensing and studies of quantum many-body spin dynamics, while coherent, near-surface defects allow for high spatial resolution imaging and coupling to other quantum elements such as phonons and photons.
Surface quality and morphology are also important in maintaining high quality factors of the diamond photonic and phononic cavities utilized in spin-phonon\cite{Lee_2017} and spin-photon coupling schemes.  \cite{mitchell2016diamondoptomechanics,khanaliloo2015nanobeamwaveguide,burek2016diamondoptomechanicalcrystals,cady2019optomechanicsNV} 

 Nitrogen-vacancy (NV) centers in diamond are presently the most widely studied diamond qubit, featuring a broad range of applications in quantum sensing, networks, and computing.
 Natural occurrence and ion implantation are two common ways to realize NV centers; however in recent years, the \emph{in situ} doping of nitrogen during plasma-enhanced chemical vapor deposition (CVD) diamond growth has become a key technique in the field.\cite{ishikawa2012optical,ohno2012engineering, Achard2019CVDandNVreview}  
 Homoepitaxial CVD growth produces a high-quality crystalline lattice with high isotopic and chemical purity, thus creating a low magnetic noise environment for preserving qubit coherence.
 Further, the gentle, bottom-up incorporation of nitrogen into this matrix via \emph{in-situ} doping provides a means for both controlling the depth localization to the nanometer scale \cite{ohno2012engineering} and for reliably producing homogeneous, coherent NV ensembles. \cite{mclellan2016patterned}
 The technique of CVD growth is well understood, \cite{butler2009understandCVD} and previous works have explored dopant incorporation into diamond films, \cite{tokuda2016morphology111,LOBAEV20171growthconditionNincorporate,tokuda2007boronmisorient} but few studies consider defect density and localization with the requirements of quantum applications in mind.
 
 Herein we explore the CVD growth of nitrogen-doped (100) diamond films as it uniquely relates to quantum applications involving quantum defects.
 Working in the slow (< 10 nm/hr) growth rate regime, we systematically investigate and quantify the growth rate, defect properties (density, coherence, and spatial distribution), and surface quality as a function of substrate miscut. We provide a model that quantitatively describes hillock nucleation in the step-flow regime of CVD growth, shedding insight on the physics of hillock formation.
 Altogether, we report an optimal miscut angle range between $\SI{0.66}{\degree}< \theta<\SI{1.16}{\degree}$ for the purposes of engineering coherent ensembles of NV centers in diamond for quantum applications.
 
\begin{figure}
\includegraphics[scale=0.50]{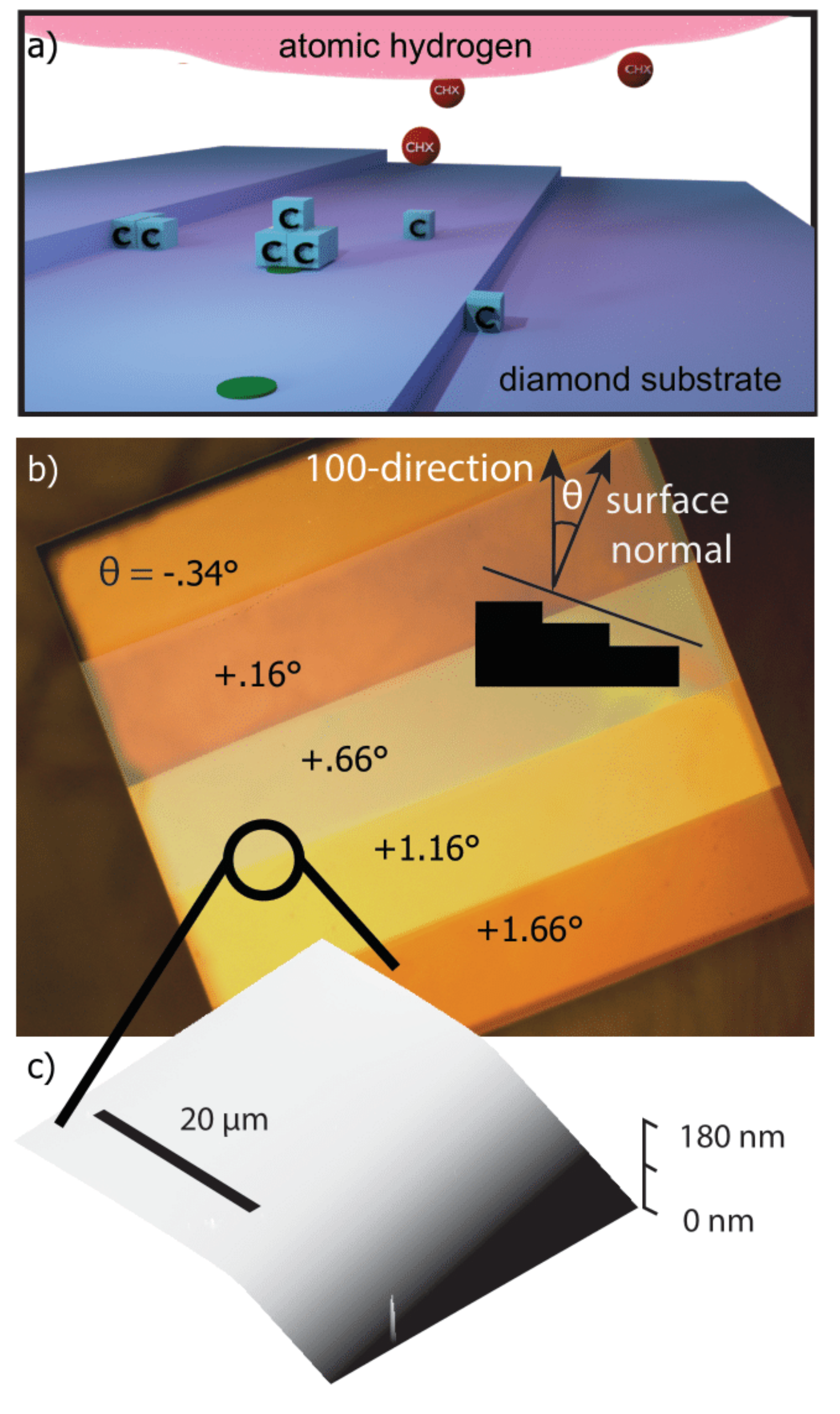}
\caption{\label{fig:intro} (a) A schematic depiction of diamond growth. The steps shown in the image are the result of the substrate miscut and are crucially important for diamond growth. Carbon incorporates preferentially at the step edges and at defects illustrated by the green circles. (b) An optical image of the surface of the diamond substrate polished at five different miscuts, with the miscut angle specified on each slice. (c) An AFM image taken near the transition between two miscut regions shown in b).}
\end{figure}

 Substrate miscut plays an important role in diamond CVD growth, as the dominant growth mechanism is thought to occur via carbon adatom incorporation at step-edge sites, \cite{butler2009understandCVD,hayashi1998epitaxialDiamond} as illustrated in the schematic of Fig.~\ref{fig:intro}a), and the miscut determines the density of step-edges on the surface.
 Furthermore, the formation of certain defects, particularly the laterally extended, flat types of growth called hillocks, occurs via a step-edge dependent mechanism. \cite{tallaire2008originOfGrowthDefects,Tokuda2015}
 These defects can both serve as a probe of the growth mode and an intriguing means of locally controlling defect density.
 
To systematically explore the influence of miscut we utilize a single sample with five regions, each polished to a different miscut angle (RMS roughness $< \SI{200}{\pico\meter}$). \cite{syntek} 
By using a single sample, we avoid sample to sample variations in substrate and sample preparation.
An optical image of this sample is shown in Fig.~\ref{fig:intro}b).
The miscut angle was varied discretely in $\SI{0.5}{\degree}$ steps across the sample, forming five different regions with miscut ranging from an absolute value of $|\theta| = \SI{0.16}{\degree}$ to $\SI{1.66}{\degree}$.
The incremental difference between regions was confirmed with atomic force microscopy (Fig.~\ref{fig:intro}c) and the miscuts were verified by X-ray diffraction (XRD), using glancing incidence reflectivity measurements to align the beam relative to the surface normal and subsequently performing a rocking curve measurement about the (004) peak.

Homoepitaxial CVD diamond growth on the multi-angle-polished substrate proceeded in the process detailed in Ref. \cite{mclellan2016patterned}.
Specifically, the growth consisted of a 3-hour undoped diamond buffer layer using 0.1 sccm of 99.999\% $^{12}$C enriched isotopically purified methane, followed by a 6-hour growth of a nitrogen doped layer using 5 sccm of 98\% $^{15}$N enriched nitrogen and 0.1 sccm isotopically purified methane.
In the last step, the sample was capped with a 4-hour undoped layer using 0.1 sccm isotopically purified methane. Throughout the growth the pressure was kept at 25 torr and the temperature was held at $\SI{800}{\celsius}$. After growth, the diamond was irradiated with $\SI{145}{\kilo eV}$ electrons (total fluence $\sim 10^{17}~\SI{}{e^{^{\small_-}}/cm^2}$) to create vacancies \cite{mclellan2016patterned} and subsequently annealed to form NV centers.
Annealing was performed in Ar/H gas at $\SI{800}{\celsius}$ for 8 hours with a 16 hour ramp time.

\begin{figure}
\includegraphics[scale=0.40]{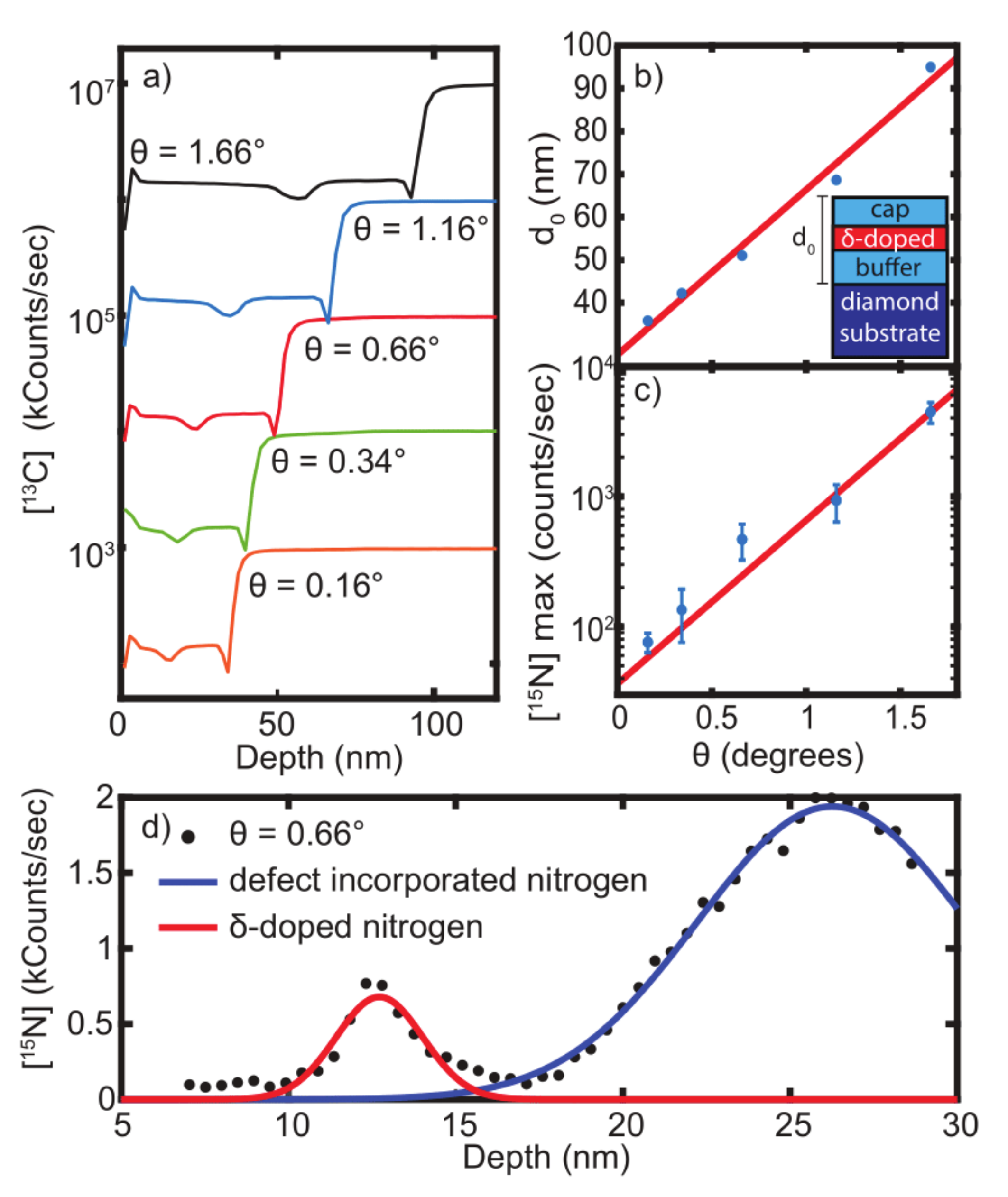}
\caption{\label{fig:nitrogen} Substrate miscut dependence of diamond growth rate and nitrogen incorporation in CVD-grown diamond. (a)$^{13}$C concentration as a function of depth as measured via SIMS for the five miscut angles studied. The depleted $^{13}$C layer indicates the grown diamond layer studied here. (b) The thickness of the grown layer, $d_0$, and (c) maximum concentration of the doped [$^{15}$N] as a function of miscut. (d) A SIMS depth profile of $^{15}$N concentration in $|\theta|=\SI{0.66}{\degree}$ miscut region. The two peaks correspond to $^{15}$N introduced via regular step-flow growth during \emph{in situ} doping and $^{15}$N present in the hillocks, respectively. The maximum value of the smaller \emph{in situ}-doped peak is plotted in (c). }
\end{figure}

We first discuss the effect of substrate miscut on the growth rate of the diamond.
Because the growth is performed with isotopically purified methane, the thickness of the grown layer, $d_0$, can be determined by measuring the thickness of the $^{13}$C-depleted layer via secondary ion mass spectrometry (SIMS).\cite{SIMS}
Figure \ref{fig:nitrogen}a) shows the $^{13}$C depth profiles in each of the five miscut regions, offset by increasing factors of 10 for ease of readability.
Each plot is averaged over several $\sim \SI{100} \times \SI{100} {\micro\meter^2}$ spots within each miscut region.
Figure \ref{fig:nitrogen}b) shows the linear increase of $d_0$ with miscut angle.
This linearity suggests that the growth proceeds by a step-flow mechanism, which depends on the density of step edges on the surface.
In this type of growth, the growth velocity, $v_g$, is related to the step velocity, $v_s$, and shallow miscut angle, $\theta$, as $v_g = v_s\sin{\theta} \approx v_s\theta$.
Thus $v_s$ can be obtained from the slope of Fig.~\ref{fig:nitrogen}b), yielding $v_s \sim\SI{100}{\pico\meter/\second}$.

Because $v_s$ appears constant over the range of miscut angles studied, it follows that the step velocity is independent of the angle, implying that the growth is dictated by the number density of available sites for carbon adatoms.
The growth mode is thus a step-limited one, rather than the adatom-limited regime that is typical in molecular beam epitaxy (MBE) growth.
We also note that for small $\theta$, the growth rate does not go to zero ($d_0(\theta=0) =\SI{28}{\nano\meter}$) indicating the dominance of a different growth mechanism in this regime.
This non-zero growth rate for a zero-miscut surface could be explained by disordered step-flow resulting from entropically stabilized steps or the presence of another growth mode arising from defects.
Lastly, we attribute the dip in the [$^{13}$C] near $d/d_0 = 0$ to be an artifact of SIMS and the dip at $d/d_0 = 1$ to be the result of the growth reaching a steady state as the plasma is ignited. The dip occurring at $d/d_0 \sim 0.5$ corresponds to the \emph{in situ} nitrogen-doped layer and likely suggests that the introduction of nitrogen changes the plasma, resulting in an altered relative isotope abundance and growth rate.

The incorporation of nitrogen during growth is also found to depend strongly on substrate miscut angle.
Figure ~\ref{fig:nitrogen}c) plots the peak $^{15}$N concentration in the \emph{in situ}-doped layer measured via SIMS as a function of miscut, showing a linearly increasing trend.
The increase of nitrogen concentration with step-edge density is consistent with the findings of Ref. \cite{Lobaev2018miscut} and suggests that the nitrogen preferentially incorporates into steps.
Examination of the SIMS $^{15}$N depth profiles in each region shows the presence of two distinct Gaussian peaks, with a representative example shown in Fig.~\ref{fig:nitrogen}d).
The first, smaller peak in the profile is found at the depth expected given the growth rate for that miscut region and the time of nitrogen introduction in the doped layer.
It is the amplitude of this peak that is plotted in Fig.~\ref{fig:nitrogen}c).
The second, broad peak is attributed to nitrogen incorporation in non-step-flow related growth features, such as hillocks or unepitaxial crystallites.
Whereas the small, \emph{in situ}-doped peak shows up at a consistent depth and amplitude across spots within the same miscut slice, the broad, defect-related peak exhibits substantial variability from spot to spot within the same miscut slice (over an area of $\sim \SI{100} \times \SI{100} {\micro\meter^2}$).
This variability is consistent with the observed variability of defect density across the sample.
Evidence for our interpretation of the larger peak and the implications of increased nitrogen incorporation will be discussed in greater detail later.

\begin{figure}
\includegraphics[scale=0.50]{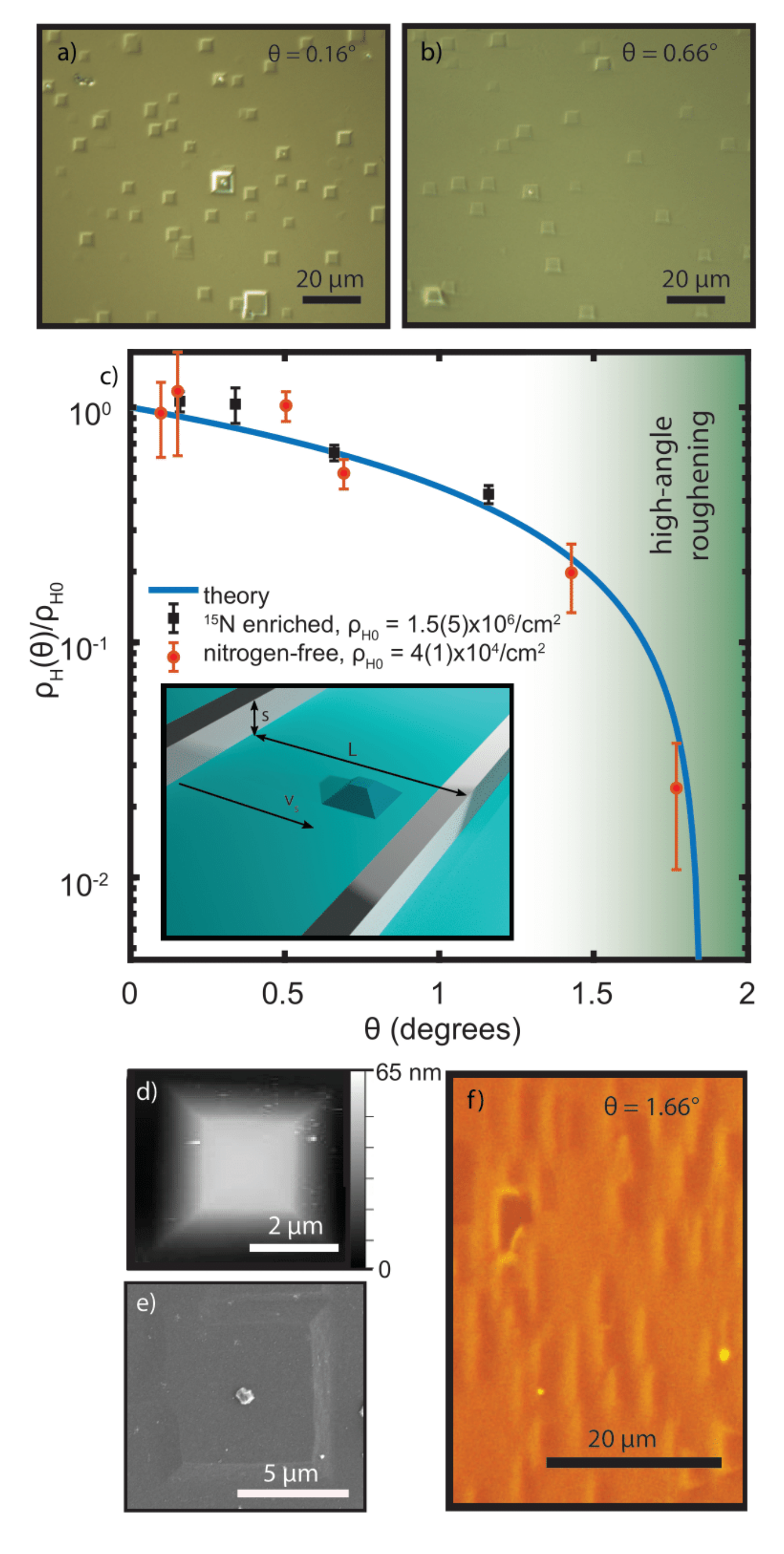}
\caption{\label{fig:hillocks} Optical images of hillocks in the (a) $\theta=\SI{0.16}{\degree}$ and (b) $\theta=\SI{0.66}{\degree}$ regions on the multi-angle sample showing increased surface defect density at higher miscuts. (c) Hillock density ($\rho_H$) normalized by $\rho_{H0}$ (hillock density at $\theta=\SI{0}{\degree}$) as a function of miscut angle for both nitrogen-doped and undoped diamond. Inset: schematic of the theoretical model for hillock formation and growth. At angles above $\gtrapprox\SI{1.66}{\degree}$, step-bunching-type growth results in hillock suppression and surface roughening via enhanced lateral growth, as seen in (f), making quantitative assignment of hillock density difficult for the $\theta=\SI{1.66}{\degree}$ region of the doped diamond sample. (d) AFM and (e) SEM images of a hillock with and without an unepitaxial crystallite in its center. (f) Optical image of the $\theta=\SI{1.66}{\degree}$ region illustrating high-angle surface roughening.}
\end{figure}

We next extend our investigation of diamond growth through selective miscut to the formation of defects such as hillocks.
Figure \ref{fig:hillocks} a-b) shows optical images of post-growth surfaces of the $\SI{0.16}{\degree}$ and $\SI{0.66}{\degree}$ regions.
The dominant feature in both images is the presence of hillocks: square, flat-topped defects that are 10's of nm tall.
An atomic force microscopy (AFM) image of a hillock defect in diamond is shown in Fig.~\ref{fig:hillocks}d).
A few of the hillocks also feature an additional unepitaxial crystallite \cite{tallaire2008originOfGrowthDefects} in their center, as shown in the scanning electron microscope (SEM) image in Fig.~\ref{fig:hillocks}e).
The density of hillock defects is found to be inversely proportional to miscut, as seen qualitatively in the images of Fig. \ref{fig:hillocks}a-b) and plotted quantitatively for all five angles in Fig. \ref{fig:hillocks}c).
For angles $\gtrapprox\SI{1.66}{\degree}$ however, the growth mode appears to transition into a step-bunching growth regime, similar to that reported in Refs. \cite{okushi2002, Shinohara1995stepbunching} and results in anisotropic defects as seen in Fig.\ref{fig:hillocks}f).
In order to preserve the well-behaved step-flow mode of growth, an optimal miscut angle less than $\sim \SI{ 1.66}{\degree}$ is desirable.

In order to understand the mechanisms of hillock growth we present a simple quantitative model of hillock nucleation and stabilization.
For shallow angles, the miscut will be related to the step dimensions as $\theta = s/L$, where $s$ is the step height and $L$ is the step length as shown in the inset of Fig.~\ref{fig:hillocks}c).
The hillocks attempt to form at sites with a density $\rho_{H0}$.
Once a hillock begins nucleating, it will continue growing and will stabilize unless an incoming step suppresses it. The time required for the hillock to reach the point of stability is defined as $T_H$, \cite{Tokuda2015} and the probability of any given hillock successfully nucleating will be given by,
\begin{equation}
    P = 1-\frac{v_sT_H}{L} = 1 - \frac{v_sT_H\theta}{s}
\end{equation}
and so the hillock density will be given by,
\begin{equation}
    \rho_H = \rho_{H0}(1 - \frac{v_sT_H\theta}{s})
\end{equation}

This model also predicts a critical angle, $\theta_C = s/v_sT_h$, at which the number of hillocks goes to zero, a result that is consistent with other observations of the hillock defect \cite{tokuda2016morphology111} and implies that at $\theta_C$ the step length is not large enough to support a hillock before lateral growth of a step edge reaches it. Fitting data in Fig. \ref{fig:hillocks}c, which also includes data for several independently grown, undoped nitrogen samples, we determine $\theta_C = \SI{1.9\pm0.2}{\degree}$.
Taking $v_s$ as determined before to be $\sim\SI{100}{\pico\meter/\second}$, the time for hillock formation is then given by, 
\begin{equation}
    T_H = \frac{s}{v_s\theta_C} \approx\SI{100}{\milli\second}
\end{equation}

 The value of $\rho_{H0}$ is found by extracting the y-intercept of the fit in Fig. \ref{fig:hillocks}c). The extracted value of $\rho_{H0}$ differs dramatically for the doped nitrogen sample ($\rho_{H0}=\SI{1.5\pm0.5e6}{\centi\meter^{-2}}$) and for the independently grown, undoped nitrogen samples ($\rho_{H0}=\SI{4\pm1e4}{\centi\meter^{-2}}$).
The undoped nitrogen growths were performed on five independent substrates with similar growth parameters and epitaxial thicknesses of 80-120 nm. Despite the radical difference in nitrogen doping, the similarity in the two curves after normalization by $\rho_{H0}$ indicates the broad applicability of the model. The two orders of magnitude difference between the hillock densities in doped and undoped samples suggests that nitrogen adatoms may be providing additional sites for hillock nucleation.
The hillock density of the undoped sample is roughly equal to the density of dislocations reported by Tokuda \emph{et al},\cite{Tokuda2015} and suggests that for undoped diamond growth, the primary source for hillock nucleation is pre-existing dislocations, aligning with the findings of Ref. \cite{tallaire2008originOfGrowthDefects}.

\begin{figure}
\begin{centering}
\includegraphics[scale=0.45]{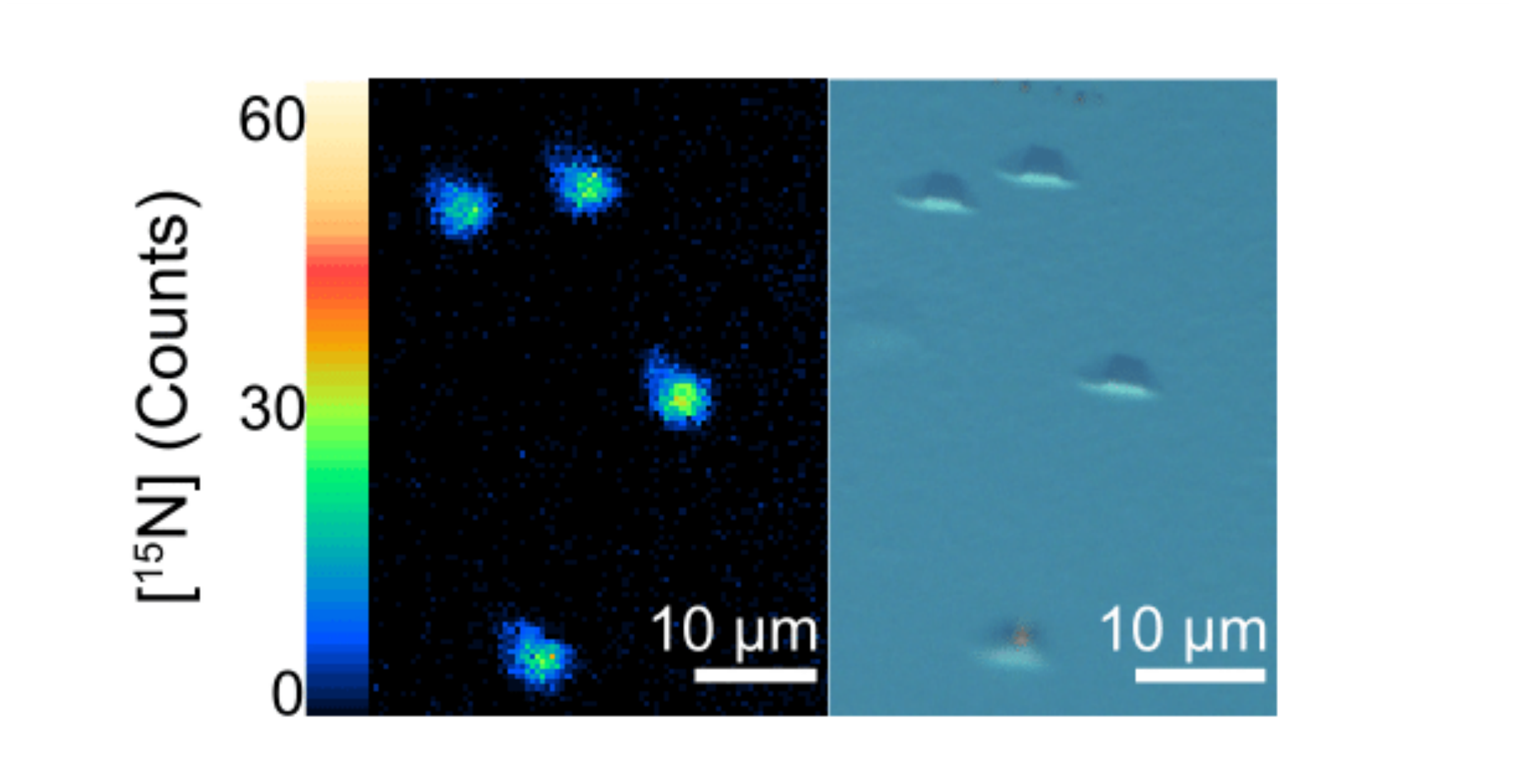}
\let\nobreakspace\relax
\caption{\label{fig:figSIMSv3} Spatially resolved SIMS image (left) (Cs$^{+}$ ion beam, $\SI{1}{\nA}$, $\SI{15}{\kilo eV}$, $\SI{2}{\micro\meter}$ spot size, $\SI{6000}{M/\Delta M}$) showing $^{15}$N enrichment at hillocks and corresponding optical image of the targeted hillocks (right).}
\end{centering}
\end{figure}

Interestingly, we find that the concentration of nitrogen is significantly enhanced at hillock sites, as shown in the spatially resolved SIMS and correlated optical images of Fig.\ref{fig:figSIMSv3}.
The spatial SIMS analysis is acquired across a $\sim \SI{100} \times \SI{100} {\micro\meter^2}$ area using a $\SI{1}{\nA}$ Cs$^{+}$ ion beam with a spot size resolution of $\SI{2}{\micro\meter}$, and the images are accumulated over a depth extent of $\sim\SI{4}{\nano\meter}$.
$^{15}$N counts within the hillocks are up to $\sim30$ larger compared to the bulk.
To confirm that the observed $^{15}$N enhancement is not a morphology-induced SIMS artifact, we also performed  spatial analysis of the $^{12}$C background and found no features in the images. 
These results further support our conclusion concerning the second, large peak observed in the $^{15}$N SIMS depth profile of Fig. \ref{fig:nitrogen}d): $^{15}$N incorporates inhomogenenously during  CVD growth and shows preference for the step-edges of the hillock defects.

\begin{figure*}
\begin{centering}
\includegraphics[scale=0.75]{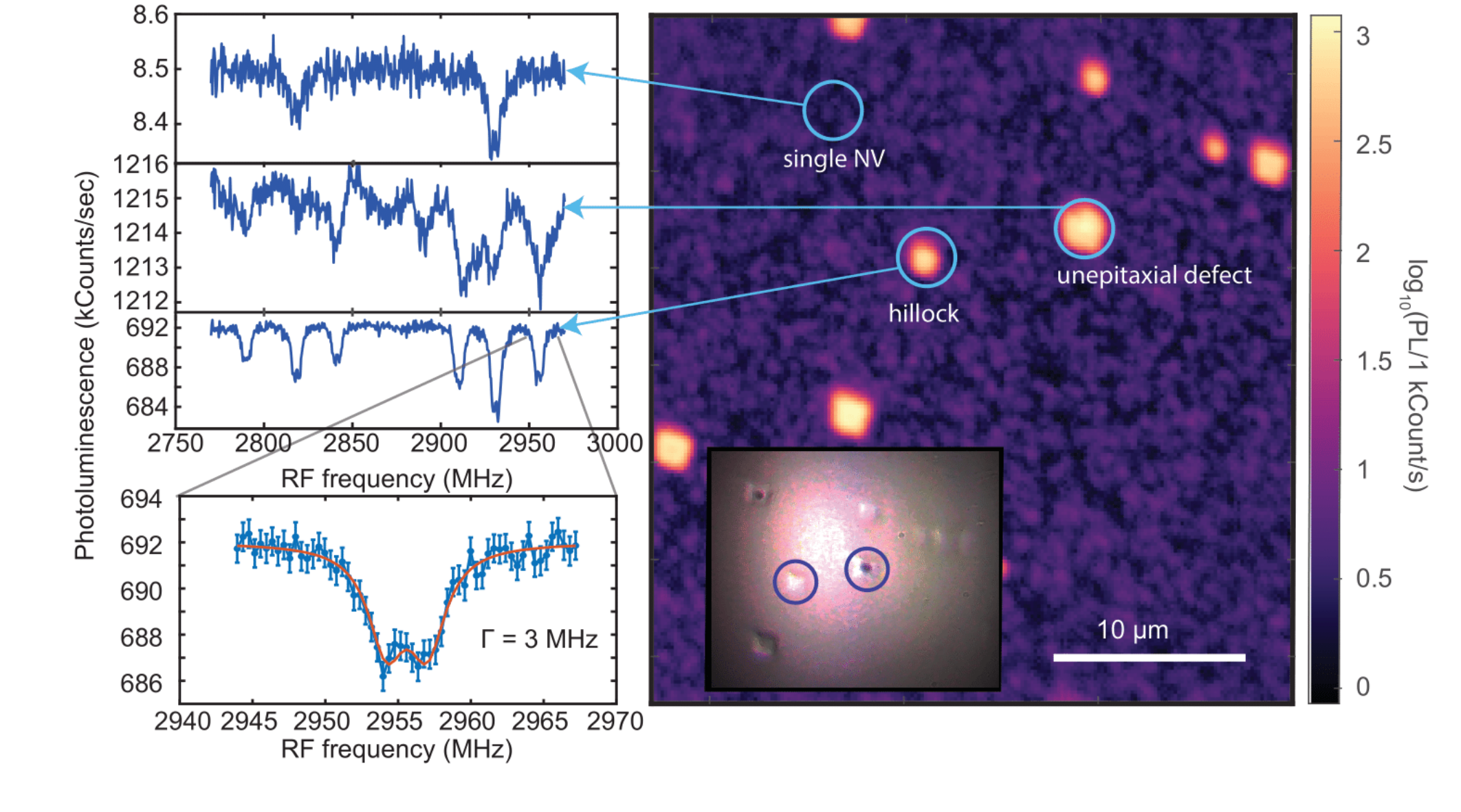}
\let\nobreakspace\relax
\caption{\label{fig:NVlinewidth} Confocal photoluminescence image (right) of the $\theta = \SI{0.66}{\degree}$ miscut region showing a single NV center as well as NVs in a hillock and unepitaxial defect. Optically detected ESR spectra (left) illustrating similar linewidths of single NV centers and those within hillock defects. Spectra taken in the unepitaxial defect locations exhibit much broader linewidths.}
\end{centering}
\end{figure*}

Lastly, we probe the spin properties of the NV centers across the sample using scanning confocal microscopy and optically detected electron spin resonance (ESR). 
Figure \ref{fig:NVlinewidth} shows a confocal image and corresponding ESR spectra in the $\theta = \SI{0.66}{\degree}$ region: single, resolvable NV centers are observed across the sample and ensembles of NV centers are seen within hillocks and unepitaxial defects.
The significantly higher NV photoluminescence signal from the defects is consistent with the expected higher NV density due to increased N incorporation.
The linewidth of the ESR peaks in the hillocks is measured to be $\Gamma = \SI{3}{\mega\hertz}$, with a resolvable $^{15}$N hyperfine splitting and is similar to the linewidth of single $^{15}$NVs in the \emph{in situ}-doped layer.
The unepitaxial defects, however, exhibit much larger linewidths and are likely strained or otherwise host an inhomogeneous material environment that results in the inhomogeneous broadening of NV centers.
These results suggest that the hillock morphology may not compromise NV quality while offering a path towards significantly higher NV density, but a more thorough analysis is necessary to draw further conclusions about NV center coherence within the hillocks.

The initial miscut of the diamond seed substrate has been shown to be important for controlling a number of parameters in CVD growth for applications in quantum science.
Namely, the growth rate, nitrogen density and hillock density are all shown to depend on the miscut angle.
Crucially, the mechanics of diamond growth change to a predominantly step bunching-dominated regime for $\theta \gtrapprox \SI{1.66}{\degree}$ and for low angles transitions to a regime of hillock-dominated growth, altogether suggesting an optimal miscut of around $\sim\SI{1}{\degree}$ for any application requiring step-flow growth.
Lastly, the inhomogeneous incorporation of nitrogen at hillock defects is reported, which opens the possibility for further investigations probing the coherence of defect-incorporated NV centers with quantum applications in mind.
Control over both lattice quality and nitrogen density are important for defect engineering in diamond and miscut provides a means for tuning both parameters.

\begin{acknowledgments}
This work was supported as part of the Center for Novel Pathways to Quantum Coherence in Materials, an Energy Frontier Research Center funded by the U.S. Department of Energy, Office of Science, Basic Energy Sciences (defect characterization and materials growth) and by the U.S. Department of Energy, Office of Basic Energy Sciences, Division of Materials Sciences and Engineering under Award No. DE-SC0019241 (materials characterization). The MRL Shared Experimental Facilities are supported by the MRSEC Program of the NSF under Award No. DMR 1720256; a member of the NSF-funded Materials Research Facilities Network. SAM acknowledges the support of the Natural Sciences and Engineering Research Council of Canada (NSERC), [funding reference number AID 516704-2018] and
the NSF Quantum Foundry through Q-AMASE-i program award DMR-1906325.

The data that support the findings of this study are available from the corresponding author upon reasonable request.
\end{acknowledgments}

% Encoding: UTF-8

%\nocite{*}
\bibliography{Miscut}{}

\begin{thebibliography}{10}

\bibitem{SIMS}
{IMS 7f-Auto}.
\newblock \url{https://www.cameca.com/products/sims/ims7f-auto}, note =
  {Accessed: 2020-03-24}.

\bibitem{Achard2019CVDandNVreview}
J.~Achard, V.~Jacques, and Alexandre Tallaire.
\newblock Cvd diamond single crystals with nv centres: a review of material
  synthesis and technology for quantum sensing applications.
\newblock 2019.

\bibitem{Awschalom2013}
David~D. Awschalom, Lee~C. Bassett, Andrew~S. Dzurak, Evelyn~L. Hu, and
  Jason~R. Petta.
\newblock Quantum spintronics: Engineering and manipulating atom-like spins in
  semiconductors.
\newblock {\em Science}, 339(6124):1174--1179, 2013.

\bibitem{awschalom2018quantum}
David~D Awschalom, Ronald Hanson, J{\"o}rg Wrachtrup, and Brian~B Zhou.
\newblock Quantum technologies with optically interfaced solid-state spins.
\newblock {\em Nature Photonics}, 12(9):516--527, 2018.

\bibitem{bluvstein2019chargeinstabilities}
Dolev Bluvstein, Zhiran Zhang, and Ania C.~Bleszynski Jayich.
\newblock Identifying and mitigating charge instabilities in shallow diamond
  nitrogen-vacancy centers.
\newblock {\em Phys. Rev. Lett.}, 122:076101, Feb 2019.

\bibitem{burek2016diamondoptomechanicalcrystals}
Michael~J. Burek, Justin~D. Cohen, Se\'{a}n~M. Meenehan, Nayera El-Sawah,
  Cleaven Chia, Thibaud Ruelle, Srujan Meesala, Jake Rochman, Haig~A. Atikian,
  Matthew Markham, Daniel~J. Twitchen, Mikhail~D. Lukin, Oskar Painter, and
  Marko Lon\v{c}ar.
\newblock Diamond optomechanical crystals.
\newblock {\em Optica}, 3(12):1404--1411, Dec 2016.

\bibitem{butler2009understandCVD}
J.E. Butler, Y.A. Mankelevich, A.~Cheesman, J.~Ma, and M.N.R. Ashfold.
\newblock Understanding the chemical vapor deposition of diamond: recent
  progress.
\newblock {\em J. Phys.: Condens. Matter}, 21(364201):1--20, 2009.

\bibitem{cady2019optomechanicsNV}
Jeffrey~V Cady, Ohad Michel, Kenneth~W Lee, Rishi~N Patel, Christopher~J
  Sarabalis, Amir~H Safavi-Naeini, and Ania C~Bleszynski Jayich.
\newblock Diamond optomechanical crystals with embedded nitrogen-vacancy
  centers.
\newblock {\em Quantum Science and Technology}, 4(2):024009, mar 2019.

\bibitem{chu2014}
Y.~Chu, N.P. de~Leon, B.J. Shields, B.~Hausmann, R.~Evans, E.~Togan, M.~J.
  Burek, M.~Markham, A.~Stacey, A.S. Zibrov, A.~Yacoby, D.J. Twitchen,
  M.~Loncar, H.~Park, P.~Maletinsky, and M.D. Lukin.
\newblock Coherent optical transitions in implanted nitrogen vacancy centers.
\newblock {\em Nano Letters}, 14(4):1982--1986, 2014.
\newblock PMID: 24588353.

\bibitem{chu2015quantum}
Yiwen Chu and Mikhail~D. Lukin.
\newblock Quantum optics with nitrogen-vacancy centers in diamond, 2015.

\bibitem{ethiermajcher2017}
G.~\'Ethier-Majcher, D.~Gangloff, R.~Stockill, E.~Clarke, M.~Hugues,
  C.~Le~Gall, and M.~Atat\"ure.
\newblock Improving a solid-state qubit through an engineered mesoscopic
  environment.
\newblock {\em Phys. Rev. Lett.}, 119:130503, Sep 2017.

\bibitem{hayashi1998epitaxialDiamond}
Kazushi Hayashi, Sadanori Yamanaka, Hideyuki Watanabe, Takashi Sekiguchi,
  Hideyo Okushi, and Koji Kajimura.
\newblock Diamond films epitaxially grown by step-flow mode.
\newblock {\em Journal of Crystal Growth}, 183(3):338 -- 346, 1998.

\bibitem{ishikawa2012optical}
Toyofumi Ishikawa, Kai-Mei~C Fu, Charles Santori, Victor~M Acosta, Raymond~G
  Beausoleil, Hideyuki Watanabe, Shinichi Shikata, and Kohei~M Itoh.
\newblock Optical and spin coherence properties of nitrogen-vacancy centers
  placed in a 100 nm thick isotopically purified diamond layer.
\newblock {\em Nano letters}, 12(4):2083--2087, 2012.

\bibitem{khanaliloo2015nanobeamwaveguide}
Behzad Khanaliloo, Harishankar Jayakumar, Aaron~C. Hryciw, David~P. Lake,
  Hamidreza Kaviani, and Paul~E. Barclay.
\newblock Single-crystal diamond nanobeam waveguide optomechanics.
\newblock {\em Phys. Rev. X}, 5:041051, Dec 2015.

\bibitem{Lee_2017}
Donghun Lee, Kenneth~W Lee, Jeffrey~V Cady, Preeti Ovartchaiyapong, and Ania
  C~Bleszynski Jayich.
\newblock Topical review: spins and mechanics in diamond.
\newblock {\em Journal of Optics}, 19(3):033001, feb 2017.

\bibitem{LOBAEV20171growthconditionNincorporate}
M.A. Lobaev, A.M. Gorbachev, S.A. Bogdanov, A.L. Vikharev, D.B. Radishev, V.A.
  Isaev, V.V. Chernov, and M.N. Drozdov.
\newblock Influence of cvd diamond growth conditions on nitrogen incorporation.
\newblock {\em Diamond and Related Materials}, 72:1 -- 6, 2017.

\bibitem{Lobaev2018miscut}
Mikhail~A. Lobaev, Alexei~M. Gorbachev, Sergey~A. Bogdanov, Anatoly~L.
  Vikharev, Dmitry~B. Radishev, Vladimir~A. Isaev, and Mikhail~N. Drozdov.
\newblock Nv-center formation in single crystal diamond at different cvd growth
  conditions.
\newblock {\em physica status solidi (a)}, 215(22):1800205, 2018.

\bibitem{mclellan2016patterned}
Claire~A McLellan, Bryan~A Myers, Stephan Kraemer, Kenichi Ohno, David~D
  Awschalom, and Ania~C Bleszynski~Jayich.
\newblock Patterned formation of highly coherent nitrogen-vacancy centers using
  a focused electron irradiation technique.
\newblock {\em Nano letters}, 16(4):2450--2454, 2016.

\bibitem{mitchell2016diamondoptomechanics}
Matthew Mitchell, Behzad Khanaliloo, David~P. Lake, Tamiko Masuda, J.~P.
  Hadden, and Paul~E. Barclay.
\newblock Single-crystal diamond low-dissipation cavity optomechanics.
\newblock {\em Optica}, 3(9):963--970, Sep 2016.

\bibitem{myers2014surfacenoise}
B.~A. Myers, A.~Das, M.~C. Dartiailh, K.~Ohno, D.~D. Awschalom, and A.~C.
  Bleszynski~Jayich.
\newblock Probing surface noise with depth-calibrated spins in diamond.
\newblock {\em Phys. Rev. Lett.}, 113:027602, Jul 2014.

\bibitem{ohno2012engineering}
Kenichi Ohno, F~Joseph~Heremans, Lee~C Bassett, Bryan~A Myers, David~M Toyli,
  Ania~C Bleszynski~Jayich, Christopher~J Palmstr{\o}m, and David~D Awschalom.
\newblock Engineering shallow spins in diamond with nitrogen delta-doping.
\newblock {\em Applied Physics Letters}, 101(8):082413, 2012.

\bibitem{okushi2002}
H.~Okushi, H.~Watanabe, S.~Ri, S.~Yamanaka, and D.~Takeuchi.
\newblock Device-grade homoepitaxial diamond film growth.
\newblock {\em Journal of Crystal Growth}, 237-239:1269 -- 1276, 2002.
\newblock The thirteenth international conference on Crystal Growth in conj
  unction with the eleventh international conference on Vapor Growth and
  Epitaxy.

\bibitem{Shinohara1995stepbunching}
Masanori Shinohara and Naohisa Inoue.
\newblock Behavior and mechanism of step bunching during metalorganic vapor
  phase epitaxy of gaas.
\newblock {\em Applied Physics Letters}, 66(15):1936--1938, 1995.

\bibitem{sohn2018controlling}
Young-Ik Sohn, Srujan Meesala, Benjamin Pingault, Haig~A Atikian, Jeffrey
  Holzgrafe, Mustafa G{\"u}ndo{\u{g}}an, Camille Stavrakas, Megan~J Stanley,
  Alp Sipahigil, Joonhee Choi, et~al.
\newblock Controlling the coherence of a diamond spin qubit through its strain
  environment.
\newblock {\em Nature communications}, 9(1):2012, 2018.

\bibitem{Steger2012}
M.~Steger, K.~Saeedi, M.~L.~W. Thewalt, J.~J.~L. Morton, H.~Riemann, N.~V.
  Abrosimov, P.~Becker, and H.-J. Pohl.
\newblock Quantum information storage for over 180 s using donor spins in a
  28si {\textquotedblleft}semiconductor vacuum{\textquotedblright}.
\newblock {\em Science}, 336(6086):1280--1283, 2012.

\bibitem{syntek}
Syntek.
\newblock Products-1: Various industrial diamonds.

\bibitem{tallaire2008originOfGrowthDefects}
A.~Tallaire, M.~Kasu, K.~Ueda, and T.~Makimoto.
\newblock Origin of growth defects in cvd diamond epitaxial films.
\newblock {\em Diamond and Related Materials}, 17(1):60 -- 65, 2008.

\bibitem{Tokuda2015}
Norio Tokuda.
\newblock {\em Homoepitaxial Diamond Growth by Plasma-Enhanced Chemical Vapor
  Deposition}, pages 1--29.
\newblock Springer International Publishing, Cham, 2015.

\bibitem{tokuda2016morphology111}
Norio Tokuda, Masahiko Ogura, Tsubasa Matsumoto, Satoshi Yamasaki, and Takao
  Inokuma.
\newblock Influence of substrate misorientation on the surface morphology of
  homoepitaxial diamond (111) films.
\newblock {\em physica status solidi (a)}, 213(8):2051--2055, 2016.

\bibitem{tokuda2007boronmisorient}
Norio Tokuda, Hitoshi Umezawa, Kikuo Yamabe, Hideyo Okushi, and Satoshi
  Yamasaki.
\newblock Hillock-free heavily boron-doped homoepitaxial diamond films on
  misoriented (001) substrates.
\newblock {\em Japanese Journal of Applied Physics}, 46(4A):1469--1470, apr
  2007.

\bibitem{toyli2012above600K_NV}
D.~M. Toyli, D.~J. Christle, A.~Alkauskas, B.~B. Buckley, C.~G. Van~de Walle,
  and D.~D. Awschalom.
\newblock Measurement and control of single nitrogen-vacancy center spins above
  600 k.
\newblock {\em Phys. Rev. X}, 2:031001, Jul 2012.

\end{thebibliography}
\bibliographystyle{plain}
\end{document}